\documentclass[aps,twocolumn,prl]{revtex4}
\usepackage{bbm}
\usepackage{amsmath}  
\def\RR{\ensuremath{\mathbbm{R}}} 
\def\id{\ensuremath{\mathbbm{1}}} 
\def\LZR{\ensuremath{\mathrm{L}^{\!\rule[-0.5ex]{0mm}{0mm}2}(\RR)}}

\def\cE{{\cal E}}
\def\cH{{\cal H}}
\def\qed{\rule{1ex}{1ex}}
\def\tr{\mathrm{tr}}
\def\ket#1{\left| #1\right>}
\def\bra#1{\left< #1\right|}
 
\def\assign{\mathrel{\raise.095ex\hbox{:}\mkern-4.2mu=}}

\begin{document}
\title{All inseparable two-mode Gaussian continuous variable states
are distillable} 

\author{G\'eza Giedke, Lu-Ming Duan, J. Ignacio Cirac, and Peter
Zoller}   

\affiliation{Institute for Theoretical Physics, University of Innsbruck,
Technikerstr. 25, 6020 Innsbruck, Austria}

\begin{abstract}
We show that all entangled Gaussian states of two infinite dimensional
systems can be distilled to maximally entangled states in finite
dimensions. The distillation protocol involves local squeezing
operations, local homodyne measurements with ancilla systems
prepared in coherent states, and local joint measurements of the
total number operator of two copies of the state.
\end{abstract}

\pacs{PACS numbers:  03.67.Hk, 03.67.-a, 03.65.Bz}


\maketitle

The existence of \emph{pure} entangled states of two or more systems
entails the possibility of finding new applications of Quantum
Mechanics, in particular in the fields of computation and
communication. In the latter case, several protocols were devised that
use entangled pairs of qubits (quantum systems described in terms of a
Hilbert space of dimension two) to achieve secure secret communication
\cite{entqubitcrypt}. These protocols have been
successfully  
implemented experimentally by using pairs of photons generated via
parametric down conversion \cite{entqubitcryptExp}. More recently,
these protocols have been generalized to the case of infinite
dimensional Hilbert spaces, i.e., the so called continuous variable
(CV) systems \cite{CVQCrypt}. The successful experimental realization
of teleportation \cite{CVQTelepTh,CVQTelepExp} using light beams in
squeezed states indicates that 
those (or more efficient) protocols may be very soon implemented in
the laboratory.  It is clear that with light beams in squeezed states
(containing many photons) one could increase the transmission rate in
quantum communication compared to the systems realized today. 

In practice, however, systems are exposed to interactions with the
environment. These interactions transform pure into \emph{mixed}
states, which may not be useful for quantum
communication. Fortunately, there exist methods to recover pure
entangled states from mixed ones in certain situations. These
processes are called \emph{entanglement distillation} (or purification)
\cite{Ben96, longpaper}, and consist of local operations and
classical communication transforming several copies of a mixed
entangled state into one (approximately) pure entangled state which is
then useful for quantum communication. In fact, using this method in
the appropriate way one can construct quantum repeaters
\cite{qrepeater} that should allow quantum communication over
arbitrarily long distances. Unfortunately, it is not known, in
general, which mixed states $\rho$ are distillable.  At the moment we
only have necessary or sufficient criteria for distillability, but not
both. First, obviously the state $\rho$ must be inseparable
(entangled) for it to be distillable. Moreover, as shown by Horodecki
\emph{et al.} \cite{Horo}, there exists a stronger necessary
condition, namely that $\rho$ must have non-positive partial
transpose. In fact, surprisingly enough there are entangled states
which are not distillable since their density matrices remain positive
under partial transposition \cite{BE}.  Furthermore, there is evidence
that this condition is not sufficient since there seem to exist states
that have non-positive partial transpose but that nevertheless are
not distillable \cite{NPTBE}. On the other hand, a useful sufficient
criterion, the so-called reduction criterion, has been reported
\cite{RC}. It states that if there exists some vector
$|\psi\rangle$ such that
\begin{equation}\label{RC}
\bra{\psi}\tr_B\rho\otimes\id-\rho\ket{\psi}<0.
\end{equation}
then the state $\rho$ is distillable. Here, tr$_B$ stands for the partial
trace with respect to the second subsystem. An important aspect of
this criterion is that if one can find a state $|\psi\rangle$
satisfying (\ref{RC}), then one can explicitly construct a protocol to
distill $\rho$.

Although all these results can be easily generalized to CV systems, in
that case it is very complicated (if not impossible, in general) even
to determine whether a mixed state is entangled or not. However, for a
class of states comprising nearly all the states that can currently be
generated in the laboratory, a necessary \emph{and} sufficient
condition for inseparability has been derived
\cite{Duan99,Sim99} (in fact, in \cite{Sim99} it is shown that the
partial transposition condition is also sufficient). Those states are
called Gaussian states, since their density matrices can be written as
Gaussian functions of position- and momentum-like operators acting on
the two infinite-dimensional Hilbert spaces of the two subsystems. As
mentioned above, this necessary and sufficient condition for
entanglement is, in general, not sufficient for a mixed state to be
distillable.  In this Letter, however, we show that for Gaussian
states of two modes inseparability implies distillability: all
entangled Gaussian states of two modes are shown to be distillable and
thus useful for quantum communication.  Moreover, our proof, which is
based in part on the reduction criterion, provides an explicit
protocol that accomplishes distillation for all those states.

Let us consider a Gaussian state $\rho$ of two systems A and B acting
on a Hilbert space $\LZR\otimes\LZR$, e.g. two modes of the
electromagnetic field. (We will in the following often use quantum
optical terms, like ``modes'' or ``beam splitters'', because quantum
optics currently offers the most promising setting for the realization
of CV systems. The results of this paper, however, are valid for all
CV systems.)
For such systems, it is convenient to describe the
state $\rho$ by its characteristic function \cite{chfct}
\begin{equation}
\chi(\xi) = {\rm tr} [\rho D(\xi)]. 
\end{equation}
Here $\xi = (q_a,p_a,q_b,p_b)\in\RR^4$ is a real vector and 
\begin{equation}
D(\xi)=e^{-i\sum_{k=a,b} (q_k X_k + p_k P_k)},
\end{equation}
where $X_{a,b}$ and $P_{a,b}$ are operators of system A and B,
respectively, satisfying canonical commutation relations ($\hbar =1$).
The characteristic function contains all the information
about the state of the system (e.g., \cite{chfct}), that is, one can
find 
$\rho$ starting from $\chi$. Gaussian states are exactly those for
which $\chi$ is a Gaussian function of $\xi$ \cite{Manu}
\begin{equation}\label{charfct}
\chi(\xi) = e^{-\frac{1}{4}\left< \xi, M\xi\right> - \left<
d,\xi \right>}, 
\end{equation}
where the \emph{correlation matrix} $M$ is a real symmetric matrix,
the \emph{displacement} $d\in\RR^4$ a real vector, and
where $\langle \cdot,\cdot \rangle$ 
denotes the standard scalar product in $\RR^4$. Thus, a
Gaussian state is fully characterized by the correlation matrix $M$
and the displacement vector $d$. 

Any Gaussian state of two modes can be transformed into what we called
the {\em standard form}, using local unitary operations only
\cite{Duan99,Sim99}. The corresponding characteristic function has
displacement $d=0$ and the correlation matrix $M$ has the simple form
\begin{equation}\label{corrmat}
M  =  \left( \begin{array}{cc} M_A&M_{AB}\\M_{AB}^T&M_B
\end{array} \right),
\end{equation}
where\\
\parbox{9cm}{
\begin{equation}\label{stdform}
M_A = \left( \begin{array}{cc} n_a&0\\0&n_a
\end{array} \right), M_B = \left( \begin{array}{cc} n_b&0\\0&n_b
\end{array} \right), M_{AB} = \left( \begin{array}{cc} k_x&0\\0&k_p
\end{array} \right).
\end{equation}}

As mentioned in \cite{Duan99} the local unitaries needed to achieve
standard form are linear Bogoliubov transformations, i.e., generated
by Hamiltonians that are at most quadratic in the operators
$X_{a,b},P_{a,b}$. For optical fields this means
that they can be performed with beam splitters, phase shifters and
squeezers, i.e., currently available technology.

The four real parameters $(n_a,n_b,k_x,k_p)$ fully characterize a
Gaussian state up to local linear Bogoliubov transformations (LLBT).
They can be easily calculated from the four LLBT-invariants $\left|
M_{A}\right|, \left| M_{B}\right|, \left| M_{AB}\right|$, and $\left|
M\right|$ as follows:
\begin{subequations}\label{param}
\begin{equation}
n_a^2 = \left| M_{A} \right|,
n_b^2 = \left| M_{B} \right|,
k_xk_p = \left| M_{AB} \right|,
\end{equation}
\begin{equation}
(n_an_b-k_x^2)(n_an_b-k_p^2) = |M|,
\end{equation}
\end{subequations}
where $|M|$ denotes the determinant of $M$; without
loss of generality we choose $k_x\ge|k_p|$. 

Note that not all values of these parameters are allowed,
since the correlation matrix must correspond to the density operator of
a physical state. It has been shown \cite{Manu} that a given
correlation matrix $M$ corresponds to a physical state if and only if
(iff) $M> 0$ and
$-JMJ\ge M^{-1}$, where $J$ is a $4\times 4$ matrix called complex
structure, defined by $J(q_a,p_a,q_b,p_b)^T=(-p_a,q_a,-p_b,q_b)^T$. In
terms of the parameters (\ref{param}) we can reexpress these
conditions as
\begin{subequations}\label{stdphys}
\begin{eqnarray}
(n_an_b-k_x^2)(n_an_b-k_p^2)+1&\geq& n_a^2+n_b^2+2k_xk_p,\\
n_an_b-k_x^2 &\geq& 1.
\end{eqnarray}
\end{subequations}   

On the other hand, starting from results of \cite{Sim99} it is easy to 
show that a Gaussian state is entangled iff the
corresponding parameters satisfy
\begin{eqnarray}\label{insep}
(n_an_b-k_x^2)(n_an_b-k_p^2)+1 &<&n_a^2+n_b^2-2k_x k_p.
\end{eqnarray}

In the following we will prove that a Gaussian state is distillable
iff its parameters fulfill (\ref{insep}); that is, iff it is
entangled.  We will proceed as follows: first, we recall the proof of
the reduction criterion in \cite{RC} and extend the result to infinite
dimensions.  The extended criterion provides a sufficient condition
for distillability of CV states. Secondly, we show for
\emph{symmetric} Gaussian states, i.e. states for which
$n_{a}=n_{b}=n$, that the inseparability criterion (\ref{insep}) is
equivalent to the reduction criterion. So, all the symmetric
inseparable states are distillable. Finally, we show that all the
states which are not symmetric ($n_{a}\neq n_{b}$) can be brought into
a symmetric form by using local operations and without disentangling
them.

We start by reviewing the proof of the reduction criterion for
distillability and generalizing it to infinite dimensions. 
Let a density matrix $\rho$ and the pure state $\ket{\psi}=\sum_{n,m}
a_{nm}\ket{n}\otimes\ket{m}$ fulfill the criterion (\ref{RC}), where
the vectors $\ket{n}$ form an orthonormal basis of $\cH$. 
The coefficients $a_{nm}$ define a matrix $A=(a_{nm})$ satisfying
$AA^\dagger = \tr_B(\ket{\psi}\bra{\psi})$.  
Distillation of $\rho$ is divided into three steps. The first is a
filtering operation: 
The operator $AA^\dagger\otimes\id$ can be viewed as an element of a
positive-operator-valued measure (POVM), 
which defines a generalized
measurement \cite{povm}. Conditional on the measurement outcome
corresponding to $AA^\dagger\otimes\id$ we
obtain the state 
\begin{equation}\label{filtering}
\tilde\rho= A^\dagger\otimes\id\rho A\otimes\id/\tr(\rho
AA^\dagger\otimes\id), 
\end{equation}
which still satisfies (\ref{RC}) but now with 
$\ket{\psi}=\ket{\Phi^N_+}:=\frac{1}{\sqrt{N}}\sum_{k=1}^N\ket{k,k}$,
the symmetric maximally entangled state of two $N$-level systems. In
this case, (\ref{RC})
implies $\tr(\tilde\rho\ket{\Phi_+^N}\bra{\Phi_+^N})>1/N$. A state
satisfying this inequality can be distilled by a generalization of the
protocol of Ref. \cite{Ben96}, which consists of two steps:
depolarization and joint measurements. Applying an operation of the
form $U\otimes U^*$ ($U$ unitary, randomly chosen) depolarizes
$\tilde\rho$, i.e.\ transforms 
it into a mixture of the maximally entangled state $\ket{\Phi_+^N}$
(which is invariant under transformations of the form $U\otimes U^*$)
and the completely mixed state $\frac{1}{N^2}\id$; the 
overlap of $\rho$ with $\ket{\Phi_+^N}$ remains unchanged.  Taking two
entangled pairs in this depolarized form, both A and
B perform the generalized XOR gate
XOR$_N:\ket{k}\ket{l}\mapsto\ket{k}\ket{(l+k)\mathrm{mod} N}$ on their
respective systems. Then they both measure the state of their second
system in the basis $\ket{k}$. The first pair is
kept, iff they get the same result. The resulting state has a density
matrix $\rho'$, which 
has a larger overlap with the maximally entangled state
$\ket{\Phi_+^N}$ than the original $\rho$. 
Iterating the last two steps sufficiently often, the overlap
between the resulting state and $\ket{\Phi_+^N}$ approaches 1,
that is, we obtain a nearly maximally entangled state.

The reduction criterion 
was proved in \cite{RC} for finite dimensional systems. Since we want
to apply it to CV states we have to show that it extends to infinite
dimensions, i.e., that Ineq. (\ref{RC}) implies distillability even
for dim$\cH=\infty$. Let $\{\ket{k}:k=0,1,...\}$ be an orthonormal
basis of $\cH$, let
$\cH_n=\mathrm{span}\{\ket{0},\ket{1},...,\ket{n}\}$, and let $\rho$ 
be a density matrix on $\cH\otimes\cH$. Let
$\cE(\rho)=\tr_B\rho\otimes\id-\rho$ be the map occuring on the
lefthand side of (\ref{RC}). Assume that
$\exists\ket{\psi}\in\cH,\epsilon>0$ such that
$\bra{\psi}\cE(\rho)\ket{\psi}<-\epsilon<0$. Since
$\rho_n=P_{\cH_n}\rho P_{\cH_n}$ converges weakly to $\rho$, there is
$N\geq0$ such that $\bra{\psi}\cE(\rho_n)\ket{\psi}<-\epsilon/2$ for
all $n\geq N$. Thus $\rho$ can be projected by local operations to a
distillable state $\rho_N$ and is therefore itself
distillable.\hfill\qed

Now we specialize to Gaussian states. 
If both the states $\rho$ and $\psi$ occuring in (\ref{RC}) are
Gaussian with displacements $d_\rho=d_\psi=0$  and correlation
matrices $M_\rho,M_\psi$ 
respectively, then (\ref{RC}) takes the form \cite{Scu98} 
\begin{equation}
2\left[ \left| M_{\tr_B\rho}+M_{\tr_B\psi} \right|
\right]^{-1/2}-4\left[ \left| M_\rho+M_\psi \right|
\right]^{-1/2}<0. 
\end{equation} 

Let $\rho$ be a symmetric Gaussian state in standard form and $\psi$
the pure two-mode squeezed state $\ket{\psi}=\frac{1}{\cosh
r}\sum_n\tanh^nr\ket{nn}$. This is also a Gaussian state and the four
parameters (\ref{param}) are $n_a=n_b=\cosh r, k_x=-k_p=\sinh r$. In
the limit of large $r$ (keeping only the leading terms in $e^r$) this
becomes after some simple algebra
\begin{equation}\label{RCsymm}
(n+k_x)(n-k_p)>1.
\end{equation}

For symmetric states, however, Ineq.\ (\ref{RCsymm}) is implied by the
inseparability criterion: for symmetric states, (\ref{insep})
simplifies to $|n^2-k_xk_p-1|<n(k_x-k_p)$, which implies
Ineq. (\ref{RCsymm}) both if $n^2-k_xk_p>1$ or $<1$, proving that all
symmetric inseparable Gaussian states are distillable.

If the state is not symmetric, it means that the reduced state at one
of the two sides has larger entropy than the other. This suggests
to let a pure state interact with the ``hotter'' side to cool it
down. To do this without destroying the entanglement of $\rho$, we
proceed as follows: $\rho$ is transformed such that the correlation
matrix of its \emph{Wigner function} takes on its standard form with
parameters $(N_a,N_b,K_x,K_p)$.  The Wigner function is the symplectic
Fourier transform of the characteristic function (\ref{charfct}) and
therefore also Gaussian for Gaussian states. The
\emph{Wigner correlation matrix} $M_W$ is related to the
(characteristic) correlation matrix by $M_W =
-JM^{-1}J$ \cite{Scu89}. One can formulate the conditions
(\ref{stdphys},\ref{insep}) similarly in terms of the parameters
$(N_a,N_b,K_x,K_p)$. Just as the form (\ref{stdform}), 
the Wigner standard form can be obtained by local squeezing
operations. 
 
Now assume that $N_b<N_a$, i.e., B is the hotter side. Take an ancilla
mode in the vacuum state and couple it to B's mode by a beam splitter
with transmittivity $\cos^2\theta$.  After a homodyne measurement of
the ancilla results a state $\tilde\rho$ with Wigner
correlation matrix $\tilde M$ with
\[\tilde M_A = \frac{1}{\nu}\left( \begin{array}{cc}
c^2N_a+s^2D_x & 
0\\ 0&c^2N_a+s^2N_aN_b
\end{array} \right),\] 
\[\tilde M_B = \frac{1}{\nu}\left(
\begin{array}{cc} N_b&0\\ 0&(c^2N_b+s^2)\nu \end{array}\right),\] 
\[\tilde
M_{AB} = \frac{1}{\nu}\left( \begin{array}{cc} cK_x&0\\0&cK_p\nu
\end{array} \right), \]
where the abbreviations $c=\cos\theta,
s=\sin\theta, \nu = s^2N_b+c^2$, and $D_{x,p}=N_aN_b-K_{x,p}^2$ were
used. The condition for symmetry, 
$|\tilde M_A| = |\tilde M_B|$, requires
\begin{equation}\label{theta}
\tan^2\theta=\frac{N_a^2-N_b^2}{N_b-D_xN_a}.
\end{equation}
Checking (\ref{insep}) for $\tilde M$ one easily sees that the sign of
the lefthand side does not change; therefore the transformed state is
inseparable iff the original one was inseparable. It remains to be
shown that there always exists a $\theta$ to satisfy (\ref{theta}),
i.e., that the right hand side of Eq. (\ref{theta}) is positive. The
numerator is positive since we have chosen $N_b<N_a$, the denominator
is positive for all states since the second part of condition
(\ref{stdphys}) implies that $(N_a-D_xN_b)>0$ and the first part
assures that $(N_a-D_xN_b)(N_b-D_pN_a)\geq(N_aK_x+N_bK_p)^2\geq0$,
hence all Gaussian states can be symmetrized this way. 
But since every Gaussian state can be brought into Wigner standard
form by local unitaries, this completes the proof.\hfill\qed

Moreover, this implies that the protocol of Ref. \cite{RC} can in
principle be used to obtain maximally entangled states in a finite
dimensional Hilbert space from any given inseparable Gaussian two-mode
state.  However, a \emph{practical} distillation protocol for all
Gaussian states remains to be found. The realistic proposals for
Gaussian states have only been shown to work for pure states
\cite{Opa99} or in the limit of small or one-sided noise
\cite{Duan99b}.  It is worth noting, though, that major parts of the
universal protocol of \cite{RC} are in fact implemented by the
procedure of Duan \emph{et al.} \cite{Duan99b}. Let us briefly discuss
the steps for distillation of an inseparable  Gaussian
state $\rho$: 

(i) remove displacements $d\not=0$; symmetrize $\rho$ if
necessary; bring the symmetric state into 
standard form. These steps require the local use of beam splitters,
one-mode squeezers, ancilla systems in coherent states, and a
homodyne measurement.\\ 
(ii) for a state in symmetric standard form the filtering operation
(\ref{filtering}) is unnecessary, since then $\rho$ already satisfies
Ineq. (\ref{RC}) with state $\ket{\psi} \propto
\lim_{\lambda\to1}\sum_k\lambda^k\ket{k}\ket{k}$ (in the photon number 
basis). This gives
$a_{nm}=\lim_{\lambda\to1}\lambda^{n+m}\delta_{nm}$, hence $A=\id$.\\
(iii) depolarize the state; in \cite{Duan99} this step is not
necessary; we do not know how to completely depolarize an arbitrary
state quantum optically. However, since all this operation does is to
increase our ignorance of the system it is not necessary for
distillation but could for any given $\rho$ be replaced by some
specific unitary that would work just as well or even better.\\  
(iv) This is the most important step of the distillation protocol. A
bilocal XOR is used to mutually entangle two entangled pairs. A
subsequent measurement selects a distilled subensemble. This operation
is implemented in \cite{Duan99b} by the total photon number
measurement at both sides: The state conditional on both A and B
obtaining the same result $N$ differs only by a local unitary
transformation (namely $\ket{n,N-n}\mapsto\ket{n,N}$) from the one
obtained by first projecting bilocally to the $N+1$ dimensional
subspace $\cH_{N+1}$ ($\rho\to\rho_{N+1}$), performing the bilocal
XOR$_{N+1}$ and finally measuring the target system.  As shown before,
for a sufficiently large value of $N$, the truncated state
$\rho_{N+1}$ is distillable and then step (iv) leads to a state with
larger overlap with $\ket{\Phi^{N+1}_+}$.\\ Iteration of the steps
(i)-(iv) leads with finite probability arbitrarily close to a
maximally entangled state in some finite dimensional space.

In conclusion we have shown that all inseparable Gaussian states of
two modes are distillable into maximally entangled states in some
finite dimensional Hilbert space. We have described a
protocol that accomplishes this and discussed its
relation to the quantum optical distillation procedure of Ref.\
\cite{Duan99b}, showing that major parts of the
protocol can be implemented by simple quantum optical means.
A number of open questions remain: firstly, it is not clear how to
implement the depolarizing operation quantum optically. 
It will be further investigated whether this step
is necessary at all for Gaussian states and how it might be performed
quantum optically. To iterate the protocol, the same question has to
be answered for the states obtained by (repeatedly) performing steps
(i)-(iv) on a Gaussian state.
Note also that the protocol \cite{Duan99b} works better, when the total
photon number measurement is performed on more than two entangled
pairs simultaneously (indeed being optimal for pure states
asymptotically). If using many modes simultaneously improves
the protocol for mixed states, too, remains to be seen. 
Furthermore, it would be interesting to see whether the equivalence of 
inseparability and distillability also holds for Gaussian states of
$m_A+m_B>2$ modes or whether one may find bound entangled Gaussian
states of bipartite systems of three or more modes. 

\begin{acknowledgments}
G.G. acknowledges financial support by the Friedrich-Naumann-Stiftung.
This work was supported by the Austrian Science Foundation under the SFB
``Control and Measurement of Coherent Quantum Systems'' (Project 11), the
European Union under the TMR network ERB--FMRX--CT96--0087 and the
project EQUIP (contract IST-1999-11053), the European
Science Foundation, and the Institute for Quantum Information GmbH,
Innsbruck. 
\end{acknowledgments}

\end{document}